\documentclass[twocolumn]{article}
\usepackage{authblk}
\usepackage{graphicx}
\usepackage{dcolumn}
\usepackage{bm}
\usepackage{float}
\usepackage{bm}
\usepackage{multicol}
\usepackage[top=0.7in, bottom=1.25in, left=1in, right=0.7in]{geometry}
\usepackage{lipsum}
\begin{document}

\title{Non-contact XUV metrology of Ru/$B_4C$ multilayer optics by means of Hartmann wavefront analysis}
\author[1]{Mabel Ruiz-Lopez*}
\author[2]{Hugo Dacasa}
\author[2]{Benoit Mahieu}
\author[2]{Magali Lozano}
\author[2]{Lu Li}
\author[2]{Philippe Zeitoun}
\author[1]{Davide Bleiner}

\affil[1]{Empa - Swiss Federal Laboratories for Materials \& Technology, \"Uberlandstrasse 129, D\"ubendorf (Switzerland)}
\affil[2]{LOA - Laboratoire d'Optiques Apliqu\'ee,  828 Boulevard des Mar\'echaux, Palaiseau (France)}
\affil[*]{Now at Desy, Notkestrasse 85, Hamburg (Germany), mabel.ruiz-lopez@desy.de}

\setcounter{Maxaffil}{0}
\renewcommand\Affilfont{\normalsize\normalfont}

\twocolumn[
  \begin{@twocolumnfalse}
  \maketitle
  \begin{abstract}
Short-wavelength imaging, spectroscopy and lithography scale-down the characteristic length-scale to nanometers. This poses tight constraints on the optics finishing tolerances, which is often difficult to characterize. Indeed, even a tiny surface defect degrades the reflectivity and the spatial projection of such optics. In this study, we demonstrate experimentally that a Hartmann wavefront sensor for extreme ultraviolet wavelengths is an effective non-contact analytical method for inspecting the surface of multilayer optics. The experiment was carried out in a tabletop laboratory using a high order harmonic generation as extreme ultraviolet source. The wavefront sensor was used to measure the wavefront-errors after the reflection of the extreme ultraviolet beam on a spherical Ru/B$_4$C multilayer mirror, scanning a large surface of approximately 40 mm in diameter. The results showed that the technique detects the aberrations in the nanometer range.
\end{abstract}
 \end{@twocolumnfalse}
]

\section{Introduction}

Applications benefiting from the extreme ultraviolet (XUV) spectral range, such as imaging at nano-scale \cite{brewer2008single, brizuela2009microscopy, ruiz_implementing}, next generation lithography\cite{takase2010imaging, balmer}, or spectroscopy \cite{peth2008near, ruiz_magentooptics, Yuni-spectroscopy}, all  need normal incidence optics based on multilayer coatings \cite{soufli2001multilayer} which demand high-quality finishing tolerances.

Minimal surface errors can be classified into three main groups named as high, mid and low-spatial-frequency errors. The high-spatial-frequency errors yield large-angle scattering of the light, reducing the energy throughput in the optical system. The mid-spatial-frequency errors produce degradation of the contrast and the scattering of the light. And the low-spatial-frequency errors quantify the distortion in the wavefronts. They are also known as \textit{figure errors} and typically characterized by the Zernike polynomials. The measurement of the latter, the low-spatial-frequency errors, is required to evaluate the image quality of the optics since those defects affect the spatial resolution significantly \cite{barty, Harris-Jones_AFM}. The tolerances are classically given by the Mar\'echal or Rayleigh criterion. As stated, the former is used when the aberrations are given in terms of their root-mean-square (RMS). According to the Mar\'echal criterion, the RMS is limited to $\lambda/14$ \cite{gautier2008optimization}. The latter, the Rayleigh lambda-quarter rule is commonly used for a given peak-to-valley (PV) wavefront error \cite{Yumoto_Rayleigh}.

A number of surface analytical metrologies are available for micro- and nano-analysis, with complementary strengths and weaknesses \cite{armelao}. All of them require a physical intrusion into the sample, with potential risk of artifacts. Most of these metrologies can be summarized into two main groups: i) those which use Atomic Force Microscopy (AFM)\cite{kozlowski1992surface} and ii) those which use interferometric techniques \cite{sommargren1996phase, weitkamp2005x}. A summary with the parameters that define both techniques is listed in Table \ref{tab:AFMvsIT}. AFMs have provided the ability to characterize the surface morphology of coated mirrors with sub-nanometer resolution  \cite{staggs1992situ}. A few groups have taken advantage of this technique for inspecting the damage that x-ray and soft x-ray sources produced on the surface of the mirrors \cite{Bleiner_multilayer, kozlowski1992surface}. Unfortunately, the surface that can be inspected is rather limited.

On the other hand, interferometric and coherent diffraction techniques can carry out \emph{non-contact} high precision metrology of wide fields-of-view \cite{yumoto2016stitching, poyneer2016x}. The most used interferometric technique in the X-ray range is the grating interferometry \cite{matsuyama2012wavefront, weitkamp2005x}, mostly used at the beamlines. Grating interferometry requires sources providing high brightness and fast repetition rates \cite{matsuyama2012wavefront, hilbert2013extreme}, only possible at large-scale facilities \cite{zhu_synchro, ichihara_synchro}. Hence, there is no routinely accessible method for such important metrology in a tabletop setup.

\begin{table}[ht]
\centering
\caption{\bf Comparison of main parameters of the atomic force microscope (AFM) and the interferometry techniques for surface metrology of XUV optics.}
\label{tab:AFMvsIT}
\begin{tabular}{p{1.8cm} p{2.6cm} p{2.6cm}}
\hline
Parameters 		& AFM \cite{george2010extreme}	                     & Interferometry techniques \cite{yumoto2016stitching,weitkamp2005x} \\
\hline
Resolution      &0.25 nm                                             & $<$1 nm   \\
Range           &$\mu$-scale                                         & mm-scale   \\
Advantage       & Observation in atmospheric conditions              & Possibility to measure wide fields-of-view and different shapes \\
Disadvantage    & Inability to measure large samples                 & Only at large facilities (FEL or synchrotron) \\
\hline
\end{tabular}
\end{table}

High-order harmonic generation (HHG) sources have emerged as a useful tool to produce coherent ultra-short-pulses of XUV radiation. To generate short-wavelengths an intense laser beam is focused on a cell containing Noble gases, i.e, Xe, Ne, Ar, etc. This results in a harmonic beam whose frequency spectrum is an odd integer multiple of the driving laser frequency. The new beam is spatially and temporally coherent, it can deliver an intensity up to $\sim$ 10$^{13-14}$ W/cm$^2$ \cite{kim2004high,rudawski2013high} and it provides a regular wavefront. With these properties, HHG shows the potential for imaging \cite{sandberg2007lensless}, interferometry  \cite{laban2014extreme} or diffraction-based measurements \cite{mercere_hartmann}.

The Hartmann technology, used to calculate the wavefronts, is based on  diffraction \cite{malacara2007optical}. Hartmann Wavefront Sensor has been used in the past for the characterization of the beam profile of free electron lasers \cite{mehrjoo2017single} and tabletop sources \cite{mercere_hartmann}. Nonetheless, the potential of the XUV Hartmann wavefront sensor has not yet been exploited for characterizing multilayer coatings. The main reason for that might be that one needs short-wavelength beams with high divergence to illuminate the full sample, which results in a reduction of the deposited fluence. We have solved this issue by keeping the initial divergence of our high harmonic generation source and by scanning the sample. The concept is similar to the stitching interferometry realized by \textit{Yumoto et at} \cite{yumoto2016stitching}. Indeed the equivalent system in the visible range, the Hartmann-Shack (using a microlens-array instead of a holes-array) was proposed for metrology of the optics surface \cite{dorn2015shack}. However, using an XUV Hartmann wavefront sensor (HWS) for wavefront metrology of the mirror surface has the following advantages: i) it works \textit{at-wavelength} (and therefore with nano-resolution), ii) it is compatible with tabletop XUV sources, iii) it is a non-contact and non-destructive technique and iv) any kind of XUV optics can be inspected.

The aim of this work was to demonstrate that an XUV Hartmann wavefront sensor is a competitive method for surface metrology of multilayer optics in a laboratory-scale setup. The paper is organized as follows: In section \ref{methods} we describe the methods used for this experiment. It includes three subsections where the XUV Hartmann wavefront sensor, the sample and the experimental setup are detailed. In Section \ref{results} we show the results obtained and we discuss them. The last section of the paper presents a summary of this technique and the main conclusions of our measurements.

 \section{Materials and methods}
 \label{methods}
 \subsection{Hartmann Wavefront Sensor}
 The wavefront measurement of the multilayer mirror was made using an XUV Hartmann wavefront sensor developed at Laboratoire d'Optique Apliqu\'ee in collaboration with Imagine Optic and SOLEIL.

\begin{figure}[h!]
\centering
\fbox{\includegraphics[width=7.3cm]{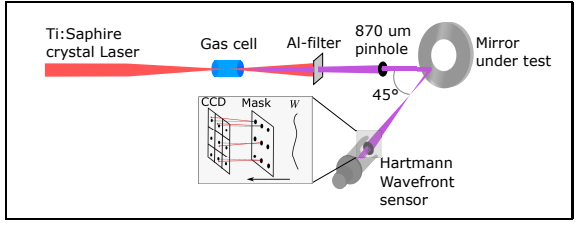}}
\caption{\label{fig:setup_wavefront}
Experimental setup. The beam was clipped by a pinhole of aperture 870 $\mu$m in order to obtain a spherical wavefront on the probe beam. The probe beam illuminates a spot in the tested mirror of approximately  $\bm{\oslash}$=1.34mm. After the mirror, the radiation wavefront conforms its shape and illuminates the mask. The new wavefront is magnified 10-folds. Finally, the sensor measures the shape of the incoming wavefront (W).}
\end{figure}

 The wavefront sensor is made up of the following two components: i) the so called Hartmann plate, formed by an array of holes distributed uniformly over its surface and ii) the CCD camera. Both are separated 210 mm apart each other. At such distance and using the adequate hole size, the plate's holes produce diffraction. The Hartmann's plate is a square metal plate which contains $65\times65$ square aperture-array of $80\times80$ $\mu$m in size \cite{mercere_hartmann,gautier2008optimization}. Each of these apertures, when the Hartmann plate is illuminated with an aberrated wavefront, produces a displaced spot with respect to the one produced by a previously-measured reference wavefront. The positions of the individual diffracted spots are measured and compared with the reference positions. The displacements are converted into a replica of the wavefront (See Figure~\ref{fig:setup_wavefront}). In order to minimize the interference due to diffraction from the adjacent holes, the square holes are rotated 25$^\circ$.

 The reconstruction of the wavefront was made with HASO software. The software calculates the wavefront by using two different algorithms: zonal and modal. In the zonal algorithm, the wavefront is divided into an array of independent zones. In each of these zones the wavefront is represented in terms of its local gradient, local curvature and phase.
 The modal reconstruction algorithm decomposes the wavefront as the sum of a set of orthogonal functions or modes expanding over the whole aperture. Legendre polynomials are used for square-shaped pupils and the Zernike polynomials for circular pupils. The expansion is defined by:

  \begin{equation}
  \Phi_{x,y}=\sum_{i=0}^{+\infty} a_i\cdot Z_i(x,y)
  \end{equation}

  Where $\Phi_{x,y}$ is the unknown wavefront, $Z_i(x,y)$ is the \textit{ith} mode and $a_i$ is the corresponding weight of the mode. This method determines accurately the low-frequency aberrations, such as astigmatism or spherical aberration, and quantifies them via the modes of the polynomials \cite{wang1980wave}. The use of the modal approach suits well for this application since the method describes completely the expected aberrations and expresses the error profile with enough level of accuracy.

HASO allows selecting an appropriate pupil diameter for each circular measurement. This is convenient for our measurements since each scan can be adapted to the real beam size. Table ~\ref{tab:coeff} shows the 8 first modes of Zernike polynomials. Each mode represents a type of aberration and it is described by a polynomial. All the polynomials depend on two parameters: $\theta$, which is the azimuthal angle, and $\rho$, the radial distance \cite{wang1980wave}.

\begin{table}[ht]
\caption{\bf Zernike polynomials and corresponding geometrical aberrations for the first 8 modes.}
\label{tab:coeff}
\begin{tabular}{p{1.0cm} p{2.7cm} p{2.9cm}}
\hline
 Mode 		& Polynomial 				& Type of aberration  \\
 \hline
1			&$\rho$cos$\theta$          & Tilt (Horizontal)  \\
2           &$\rho$sin$\theta$          & Tilt (Vertical)   \\
3			&2$\rho^2$-1                & Defocus (Piston)   \\
4			&$\rho^2$cos(2$\theta$)		& Astigmatism at 0$^\circ$ \\
5			&$\rho^2$sin(2$\theta$)		& Astigmatism at 45$^\circ$ \\
6           &$(3\rho^2$-2)cos(2$\theta$)& Coma at 0$^\circ$  \\
7			&$(3\rho^2$-2)sin(2$\theta$)& Coma at 45$^\circ$ \\
8			& 6$\rho^4$-6$\rho^2$+1		& 3rd order spherical aberration  \\
\hline
\end{tabular}
\end{table}

\subsection{The multilayer sample}
\label{mirror}

The sample was a spherical multilayer mirror coated with Ru/B$_4$C. The substrate was fused silica. The mirror was optimized for $\lambda=$12 nm, obtaining a reflectivity close to 50\%. The mirror under test is the primary mirror of our Schwarzschild objective described in \cite{ruiz_implementing}. A Schwarzschild objective is made of two concentric mirrors: a concave primary mirror and a convex secondary mirror. It has a theoretical spatial resolution of approximately 60 nm. The main parameters of the sample are summarized in Table ~\ref{tab:mirror}.

\begin{table}[h]
\caption{\bf Parameters of the multilayer mirror used as a sample for this experiment.}
\label{tab:mirror}
\begin{tabular}{ccc}
\hline
Parameters 		                & Value  \\
\hline
Radius of curvature			 	&100 mm \\
Diameter					    &(38.1 $\pm$ 0.1) mm    \\
			                    & Internal hole: (17.01 $\pm$ 0.1) mm   \\
Thickness				        &(8.00 $\pm$ 0.1) mm  \\
Reflectivity at 12nm            &(50-55)\%\\
\hline
\end{tabular}
\end{table}

\begin{figure}[h]
\centering
\fbox{\includegraphics[width=8cm]{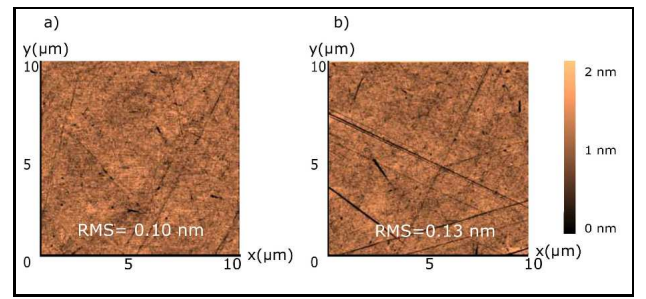}}
\caption{Atomic force microscopy of the substrate surface of the sample. Measurements realized at the center (a) and at 15 mm from the center (b). The substrate roughness is shown to be uniform for the deposition of the coating.}
\label{fig:RMS_Berlin}
\end{figure}

The deposition of the multilayer was done using an XUV sputtering system. In order to preserve the image quality, a first characterization of the mirror substrate was done at the PTB Berlin (BESSY II) as shown in Figure \ref{fig:RMS_Berlin}. The substrate roughness is shown to be uniform (RMS$<$1nm) for the deposition of the coating.

Surface characterization showed a substrate roughness of 0.10 nm (RMS) in the central part and 0.13 nm (RMS) at 15 mm from the center. The substrate is subject to the Mar\'echal criterion for the designed wavelength $\lambda$=12 nm, thus RMS = $\frac{\lambda}{14}$= 0.85 nm. Our experimental source runs at 32 nm-wavelength, which corresponds to a Mar\'echal tolerance RMS=2.3 nm.

   \subsection{XUV metrology setup}

The setup is illustrated schematically in Figure ~\ref{fig:setup_wavefront}. All the measurements were done under controlled vacuum (10$^{-5}$ mbar). The multilayer mirror under test is illuminated with a high harmonic beam with a divergence of approximately 2 mrad. They are generated by focusing an infrared (IR) beam from a Ti:Sapphire laser, with a  repetition rate of 1 kHz and pulse energy of 7 mJ, through a cylindrical gas cell with diameter 10 mm and length 15 mm, filled with Argon. The pressure in the cell was 60 mbar. An Aluminum filter was used to filter out the remaining IR radiation. The experimental setup (cell length, gas pressure, focusing distance of the IR laser, etc...) was optimized to produce a set of few harmonics (23rd at 34.8 nm, 25th at 32 nm and 27th at 29.6 nm) with an average value of 32.14 nm, taken later on to quantify or display the wavefront aberrations.  The probe beam was clipped with a pinhole handing an aperture of 870 $\mu$m, such only the central part, which has a spherical wavefront, illuminates the mirror.

The multilayer-sample was placed 235 mm after the pinhole. In order to illuminate the full sample, one needs high divergence beams, which results in a reduction of the deposited fluence. We solved this issue by utilizing the initial divergence of our HHG source and stitching the sample. For this aim the mirror was mounted on a rotating stage and the sample rotated around the normal by 3$^\circ$ between consecutive measurements. This ensured that the beam covered a full circumference around the center of the mirror, at 28 mm from it. The scan covered approximately 170 mm$^2$ of the sample's surface (120 measurements of illumination spots with diameter $\bm{\oslash}$=1.34mm).

The final position of the HWS is also determined by the divergence. In this case the imaging divergence (after the reflection) was enough to almost illuminate the full Hartmann plate. Due to its curvature, the mirror focused the beam at a plane 63 mm downstream the reflection of the light. From this focal plane on, the beam diverges with identical Numerical Aperture (NA) up to the CCD. The Hartmann plate is almost fully illuminated relatively close, at 750 mm from the mirror. The measurements made with the Hartmann wavefront sensor experience a magnification (10-folds) due to the sphericity of the tested multilayer mirror. All the results shown in this section are corrected by this effect.

\section{Results and discussion}
  \label{results}

\begin{figure}[H]
\centering
	\fbox{\includegraphics[width=7.5cm]{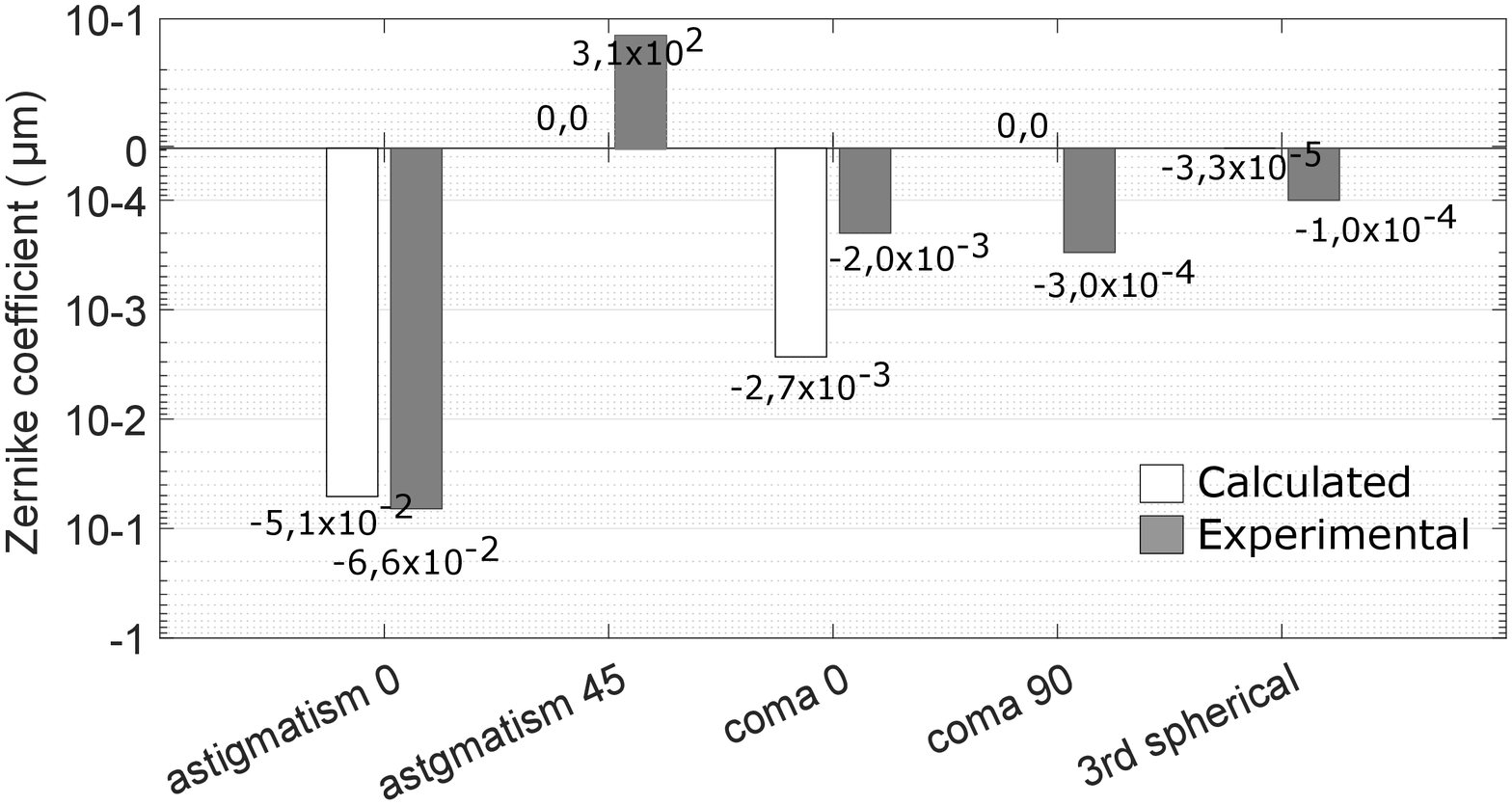}}
	\caption{Zernike coefficients for the calculated (white bars) and experimental results (gray bars).The major distortion was found for the astigmatism at 0$^\circ$.}
	\label{fig:zernike}
\end{figure}

The experimental setup was simulated with OSLO ray tracing, obtaining the values shown in Figure ~\ref{fig:zernike}. The major distortion was found for the astigmatism at 0$^\circ$. The calculated Zernike coefficient (-5.1x10$^{-2}\mu$m) and the value obtained with HASO during the experiment (-6.6x10$^{-2}\mu$m) are in the same range. The accuracy ($\varepsilon$) between both data is $\varepsilon$=78.4$\%$. For the astigmatism at 45$^\circ$, our simulation does not show any value (Zernike coefficient = 0 $\mu$m), however, there is 3.1x10$^{-2}\mu$m offset in the experimental measurements. We calculated that this offset corresponding to a tilt vertex of approximately 6$^\circ$. When this tilted mode is implemented on the simulations, we obtained a better match. The offset is probably a consequence of a misalignment of the optics in the setup during the measurements. Assuming also a misalignment at 0$^\circ$, i.e., the beam and the Hartmann are aligned at 49$^\circ$ instead of 45$^\circ$, the accuracy increases to $\varepsilon$=98.8$\%$. Other aberrations, like coma and spherical aberrations, show very low values, but also close to the simulation.

\begin{figure*}[ht]
\centering
	\fbox{\includegraphics[height=7.5cm]{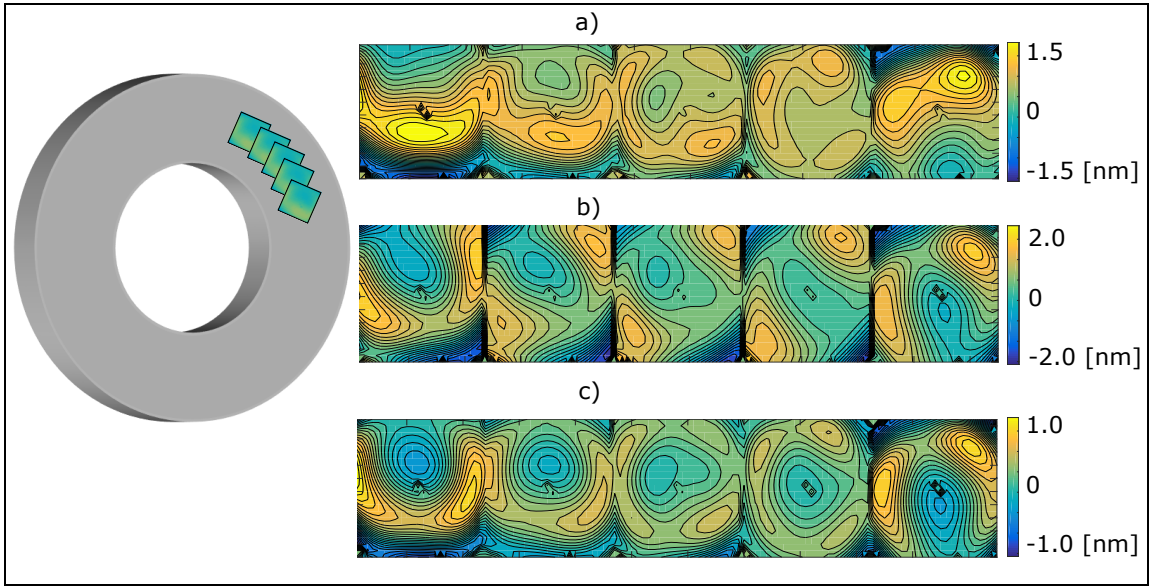}}
	\caption{Experimental wavefront for five consecutive scanned areas. a) Spherical aberration has a PV=2.2 nm, b) Astigmatism at 45$^\circ$ has a PV=2.8 nm, c) Higher order aberrations (coma, distortion and Zernike mode number higher than 6$^th$) have a PV=2.2 nm.}
	\label{fig:wavefront}
\end{figure*}

Figure ~\ref{fig:wavefront} shows in detail the experimental wavefront for five consecutive scans. The software is able to decompose the wavefront into individual aberrations. Figure ~\ref{fig:wavefront}a) corresponds to the spherical aberration. The measurements show a peak to valley PV=2.2 nm (from -1.4 nm valley to 0.8 nm peak).  Slight differences can be appreciated on each wavefront. In order to obtain exactly the same wavefront profiles one would have to observe scans at the very same distance from the center, which is almost impossible. The spherical aberration is an intrinsic defect of the multilayer mirror due to the radius of curvature.

Figure~\ref{fig:wavefront}b) shows the astigmatic aberrations. The measurements show a peak to valley PV=2.8 nm (from -1.9 nm valley to 0.9 nm peak). We only illustrate the astigmatism at 0$^\circ$, since the oblique astigmatism (45$^\circ$) is due only to the horizontal misalignment of the sample. All the measurements show very similar profiles.

Figure ~\ref{fig:wavefront}c) shows the higher order aberrations, i.e. those represented by a Zernike polynomial over the 6$^{th}$ mode:  coma, trefoil, etc. These are the aberrations that could affect the reflectivity and the spatial resolution of the mirror. The peak to valley, in the measurements, is PV=2.2 nm (from -1.3 nm valley to 0.9 nm peak).

In order to evaluate the wavefront-error, we used the Rayleigh quarter-wavelength rule $\phi=\lambda /4$= 8.0 nm, where $\phi$ is the wavefront-error and $\lambda$ is the wavelength. The higher order aberrations, were shown to have peak to valley of PV=2.2 nm, much smaller (3.6X) than the theoretical tolerance resolution due to Rayleigh rule. It is important to remember at this point, that the tested mirror is the primary mirror of a Schwarzschild objective with a resolution bellow 100 nm, and the PV obtained by this method is substantially smaller.

\section{Conclusions}%
We characterized the surface of a spherical multilayer mirror with nano-scale resolution in a tabletop laboratory by performing an XUV Hartmann wavefront sensor. The setup proposed for the experiment overcomes the difficulties such as low flux or large setup distance. The measurements showed that the wavefront deformations at the surface of the tested mirror are negligible. The PV values obtained are bellow the tolerances error indicated by the Rayleigh rule. The largest PV value found for higher order aberrations was 2.2 nm (<$\lambda/4$). Spherical and astigmatism aberrations measured on the mirror were introduced artificially in the setup due to the tilt and spherical shape of the tested multilayer mirror. Our calculations in OSLO were used to double-check our results, and we obtained Zernike values very close to the experimental results. We conclude that our method allows precise measurement of the surface of XUV-optics in the home-laboratory and it is easy to implement in other laboratories. The instrument has demonstrated to be an efficient non-contact technique for surface metrology for low brightness sources.

\section{Funding Information}

The present work was supported by the  LASERLAB-EUROPE, the \textit{Germaine de Staël} program Funding (N. dossier : 805871G) and the Swiss National Science Foundation under (Grant Number PP00P2-133564/1)
© 2018 Optical Society of America]. One print or electronic copy may be made for personal use only. Systematic reproduction and distribution, duplication of any material in this paper for a fee or for commercial purposes, or modifications of the content of this paper are prohibited.

\end{document}